\documentclass[aps,prc,twocolumn,superscriptaddress,showpacs,nofootinbib,floatfix,showkeys]{revtex4}

\usepackage{graphicx}
\usepackage{dcolumn}
\usepackage{bm}

\newcommand{\pt} {\mbox{$p_T$}}
\newcommand{\mt} {\mbox{$m_T$}}
\newcommand{\snn} {\mbox{$\sqrt{s_{NN}}$}}
\newcommand{\meanpt} {\mbox{$\langle p_{T} \rangle$}}

\newcommand{\pbar} {\mbox{$\overline{p}$}}

\newcommand{\npart} {\mbox{$N_{part}$}}
\newcommand{\npartav} {\mbox{$\langle N_{part} \rangle$}}

\def\Journal#1#2#3#4{#1 {\bf #2}, #3 (#4)}

\def\JP{J.~Phys.}

\def\NIM{Nucl. Instr. and Meth.}

\def\NP{Nucl. Phys.}

\def\PL{Phys. Lett.}

\def\PRL{Phys. Rev. Lett.}
\def\PR{Phys. Rev.}

\begin{document}
\title{Centrality Dependent Particle Production \\ 
       at $y=0$ and $y \sim1$ in Au+Au Collisions \\ 
       at $\snn$~=~200 GeV}

\newcommand{\bnl}{Brookhaven National Laboratory, Upton, New York 11973}
\newcommand{\ires}{Institut de Recherches Subatomiques and Universit{\'e} Louis Pasteur,Strasbourg, France}
\newcommand{\kraknuc}{Institute of Nuclear Physics, Krakow, Poland}
\newcommand{\krakow}{Smoluchkowski Inst. of Physics, Jagiellonian University, Krakow, Poland}
\newcommand{\baltimore}{Johns Hopkins University, Baltimore, Maryland 21218, USA}
\newcommand{\newyork}{New York University, New York, New York 10003, USA}  
\newcommand{\nbi}{Niels Bohr Institute, Blegdamsvej 17, University of Copenhagen, Copenhagen 2100, Denmark}
\newcommand{\texas}{Texas A$\&$M University, College Station, Texas, 17843, USA}   
\newcommand{\bergen}{University of Bergen, Department of Physics, Bergen, Norway}   
\newcommand{\bucharest}{University of Bucharest, Romania}
\newcommand{\kansas}{University of Kansas, Lawrence, Kansas 66045, USA}   
\newcommand{\oslo}{University of Oslo, Department of Physics, Oslo, Norway}    
\affiliation{\bnl}
\affiliation{\ires}
\affiliation{\kraknuc}
\affiliation{\krakow}
\affiliation{\baltimore}
\affiliation{\newyork}
\affiliation{\nbi}
\affiliation{\texas}
\affiliation{\bergen}
\affiliation{\bucharest}
\affiliation{\kansas}
\affiliation{\oslo}
\author{I.~Arsene}          \affiliation{\bucharest}
\author{I.~G.~Bearden}      \affiliation{\nbi} 
\author{D.~Beavis}          \affiliation{\bnl} 
\author{C.~Besliu}          \affiliation{\bucharest} 
\author{B.~Budick}          \affiliation{\newyork} 
\author{H.~B{\o}ggild}      \affiliation{\nbi} 
\author{C.~Chasman}         \affiliation{\bnl} 
\author{C.~H.~Christensen}  \affiliation{\nbi} 
\author{P.~Christiansen}    \affiliation{\nbi} 
\author{J.~Cibor}           \affiliation{\kraknuc} 
\author{R.~Debbe}           \affiliation{\bnl} 
\author{E.~Enger}           \affiliation{\oslo}  
\author{J.~J.~Gaardh{\o}je} \affiliation{\nbi} 
\author{M.~Germinario}      \affiliation{\nbi} 
\author{K.~Hagel}           \affiliation{\texas} 
\author{H.~Ito}             \affiliation{\bnl} 
\author{A.~Jipa}            \affiliation{\bucharest} 
\author{F.~Jundt}           \affiliation{\ires} 
\author{J.~I.~J{\o}rdre}    \affiliation{\bergen} 
\author{C.~E.~J{\o}rgensen} \affiliation{\nbi} 
\author{R.~Karabowicz}      \affiliation{\krakow} 
\author{E.~J.~Kim}          \affiliation{\bnl}\affiliation{\kansas}  
\author{T.~Kozik}           \affiliation{\krakow} 
\author{T.~M.~Larsen}       \affiliation{\oslo} 
\author{J.~H.~Lee}          \affiliation{\bnl} 
\author{Y.~K.~Lee}          \affiliation{\baltimore} 
\author{S.~Lindal}          \affiliation{\oslo} 
\author{R.~Lystad}          \affiliation{\bergen}
\author{G.~L{\o}vh{\o}iden} \affiliation{\oslo} 
\author{Z.~Majka}           \affiliation{\krakow} 
\author{A.~Makeev}          \affiliation{\texas} 
\author{M.~Mikelsen}        \affiliation{\oslo} 
\author{M.~Murray}          \affiliation{\texas} \affiliation{\kansas}
\author{J.~Natowitz}        \affiliation{\texas} 
\author{B.~Neumann}         \affiliation{\kansas} 
\author{B.~S.~Nielsen}      \affiliation{\nbi} 
\author{D.~Ouerdane}        \affiliation{\nbi} 
\author{R.~P\l aneta}       \affiliation{\krakow} 
\author{F.~Rami}            \affiliation{\ires} 
\author{C.~Ristea}          \affiliation{\bucharest} 
\author{O.~Ristea}          \affiliation{\bucharest} 
\author{D.~R{\"o}hrich}     \affiliation{\bergen} 
\author{B.~H.~Samset}       \affiliation{\oslo} 
\author{D.~Sandberg}        \affiliation{\nbi} 
\author{S.~J.~Sanders}      \affiliation{\kansas} 
\author{R.~A.~Scheetz}      \affiliation{\bnl} 
\author{P.~Staszel}         \affiliation{\krakow} \affiliation{\nbi} 
\author{T.~S.~Tveter}       \affiliation{\oslo} 
\author{F.~Videb{\ae}k}     \affiliation{\bnl} 
\author{R.~Wada}            \affiliation{\texas} 
\author{Z.~Yin}             \affiliation{\bergen}
\author{I.~S.~Zgura}        \affiliation{\bucharest}
\collaboration{BRAHMS Collaboration} \noaffiliation
  
\date{\today}
\begin{abstract}
Particle production of identified charged hadrons, 
$\pi^{\pm}$, $K^{\pm}$, $p$, and $\bar{p}$ in Au+Au collisions 
at $\snn =$ 200 GeV has been studied as a function
of transverse momentum and collision centrality at $y=0$ and 
$y\sim1$ by the BRAHMS experiment at RHIC. 
Significant collective transverse flow at kinetic freeze-out has 
been observed in the collisions. 
The magnitude of the flow rises with the collision centrality.  
Proton and kaon yields relative to the pion production increase
strongly as the transverse momentum increases and 
also increase with centrality. 
Particle yields per participant nucleon
show a weak dependence on the centrality for all particle species.  
Hadron production remains relatively constant within one unit 
around midrapidity in Au+Au collisions at $\snn =$ 200 GeV.
\end{abstract}
\maketitle

\section{INTRODUCTION}
\label{sec:intro}

The primary goal of the relativistic heavy-ion collider (RHIC) is to
create and study matter at extremely high energy density.
It is hypothesized that at the energy densities
reached in central Au+Au reactions at RHIC, the matter created is
composed of de-confined colored objects 
~\cite{review1,review2, review3}. 
A summary of the results and opinions of the four experimental collaborations 
on the status of achieving this goal can be found in their ``White Papers" 
\cite{BrahmsWhite,PhobosWhite,PhenixWhite,StarWhite}.
We expect that the signals of any de-confined phase 
should become stronger as the overlapped region increases in Au+Au collisions.
Testing this hypothesis requires studying particle production 
as a function of centrality.
The particle distributions in transverse momentum and rapidity 
may provide a key to understanding any non-hadronic effects 
that might appear in central nucleus-nucleus collisions.
Pions, kaons, protons and antiprotons are the most abundantly
produced particles in the high-energy heavy-ion collisions, 
and they carry information about the bulk properties 
of the nuclear matter created from the collisions. 

Pions, being the lightest of the produced hadrons are thus 
the most copiously produced, and their numbers can be related 
to the entropy density of the emitting source.
Kaons carry a significant fraction of the total strangeness produced~\cite{kaon}.
Protons and antiprotons provide an experimental tool for measuring 
baryon production and allow us to explore baryon transport 
from beam rapidity toward midrapidity~\cite{brahms_stopping, bass}. 
The global thermodynamic properties and collective motion of the system 
at the kinetic freeze-out point can be deduced, albeit, in a model dependent way, 
from transverse momentum spectra 
as a function of rapidity and centrality~\cite{ref:blast-wave}.    

In this paper, we present transverse momentum spectra, yields, and ratios
for identified charged hadrons ($\pi^{\pm}$, $K^{\pm}$, $p$, $\bar{p}$)
obtained with the BRAHMS Mid-Rapidity Spectrometer. 
We have measured these spectra for $y=0$ and $y\sim1$ 
as a function of collision centrality. 
At midrapidity, our observations are in agreement with the result of
the PHENIX experiment~\cite{phenix_mid} within systematic uncertainties.
The data presented here are available at~\cite{brdata}.

\section{EXPERIMENTAL DETAILS}
\label{sec:brahms}

The BRAHMS experiment consists of two movable magnetic spectrometer arms,
the Mid-Rapidity Spectrometer (MRS) and the Forward Spectrometer, 
and global detectors for event characterization.

\begin{figure}[htb]
\includegraphics[width=\columnwidth]{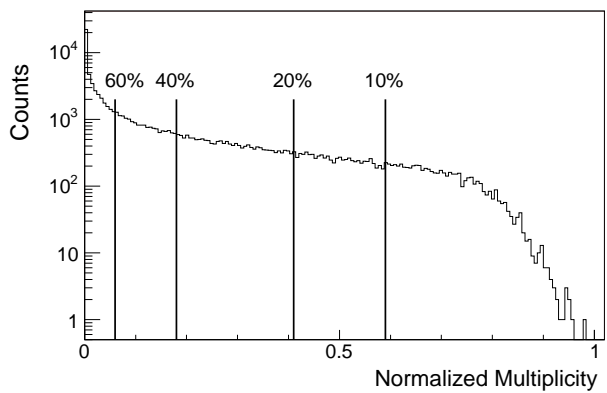}
\caption{MA array multiplicity distribution. Lines show 
the limits for indicated centralities.}
\label{fig:mult}
\end{figure}

In order to characterize the centrality of collisions,
a multiplicity array (MA) consisting of a coaxial arrangement of
Si strip detectors and scintillator tiles 
surrounding the intersection region is employed.
The Si strip detectors and scintillator tiles give independent
measurement of charged-particle multiplicities allowing the
two measurements to be averaged in the final determination.
The pseudo-rapidity coverage of the MA is approximately 
$-2.2<\eta<2.2$~\cite{200gevmult,tilenim}.
The centrality selection is obtained by developing 
minimum-bias trigger events, which are defined 
using two Zero Degree Calorimeters (ZDC), 
requiring energy deposit equivalent to at least
one neutron in each of the two detectors
and also requiring a signal in the MA 
to reject Coulomb dissociation events~\cite{200gevmult, nim}.
Figure~\ref{fig:mult} shows the charged-particle multiplicity distribution 
for minimum-bias events in the range of $-2.2<\eta<2.2$.
The lines on the plot indicate four centrality
windows in the analysis, 0$-$10\%, 10$-$20\%, 20$-$40\%, and 40$-$60\%,
where 0\% corresponds to the most central events.

The number of participating nucleons ($\npart$) in Table~\ref{tab:npart} 
are estimated using the Glauber Monte-Carlo HIJING calculation~\cite{hijing}.
More peripheral collision events were not included in this paper
because of limited statistics.  

\begin{table}[htb]
\begin{ruledtabular}
\begin{tabular}{cllc}
   & Centrality    & $\npartav$     & \\\hline
   &   $ 0-10\%$   &  328 $\pm$ 6   & \\
   &   $10-20\%$   &  239 $\pm$ 10  & \\
   &   $20-40\%$   &  140 $\pm$ 11  & \\
   &   $40-60\%$   &   62 $\pm$ 10  & \\
\end{tabular}
\end{ruledtabular}
\caption{The number of participant nucleons $\npart$ estimated
from the HIJING model~\cite{hijing}.}
\label{tab:npart}
\end{table}

The uncertainty in determining centrality from the multiplicity distribution
was estimated to be $\pm1.7\%$ for the most central bin 
and $\pm9.4\%$ for the most peripheral bin.  
The fraction of the inclusive yield lost by the minimum-bias trigger
is estimated to be about 4\% and is corrected for.   
The location of the collision vertex was determined to an accuracy of 0.7 cm
using Beam-Beam Counters (BBC)~\cite{nim}. 
The BBCs are located 2.2~m on either side 
of the nominal interaction point (IP) and also
provide a start time~(T0) for time-of-flight measurement. 

The MRS is a single-dipole-magnet spectrometer
which, by rotating about the nominal collision point, 
provides the angular coverage of $30^\circ<\theta<95^\circ$, where
$\theta$ is the polar angle with respect to the beam axis.
The MRS contains two time projection chambers~(TPCs), TPM1 and TPM2,
which determine the three dimensional trajectories 
of the charged particles through the spectrometer. 
Between the two TPCs there is a dipole magnet~(D5) for momentum determination.
This assembly is followed by a highly segmented scintillator
time-of-flight wall~(TOFW).

The BRAHMS TPCs are located at a distance 0.95~m~(TPM1) 
and 2.85~m~(TPM2) from the interaction point. 
Each TPC is a rectangular box filled with 90\% Ar and 10\% CO$_{2}$.
The ionization produced by charged particles is collected on an anode grid.
This grid is divided along the particle path into 
12 rows (TPM1) and 20 rows (TPM2). 
Each row has 96 pads (TPM1) and 144 pads (TPM2) 
transverse to the direction of a normal-incident particle.
The mapping of row, pad, and drift time leads to three-dimensional
space points. The averaged resolutions measured from track residuals
are 310$-$387~$\mu$m for X~(pad) and 427$-$490~$\mu$m for Y~(time).
Details can be found in Ref.~\cite{nim,peter,bhs,jij}.

Track reconstruction starts by finding straight-line track segments 
in the TPCs. The reconstructed straight tracks are joined 
inside the analyzing magnet by taking an effective edge approximation,
and the momentum associated with the tracks are calculated from
the vertical magnet field, the length in the magnetic field region, 
the polar angle of the tracks with respect to the matching plane,
and the averaged vertical slope of the tracks.
The matching plane is defined as the vertical plane that contains 
the perpendicular bisector of the line joining the effective edge 
entry and exit points of the tracks.
Once the momentum is known
the reconstructed tracks are projected toward the beam axis and
checked for consistency with the collision vertex determined using the BBCs.
For this analysis we only use tracks that project to within $\pm$12.5~cm  
of the nominal vertex in the horizontal plane.

\begin{figure}[htb]
\includegraphics[width=\columnwidth]{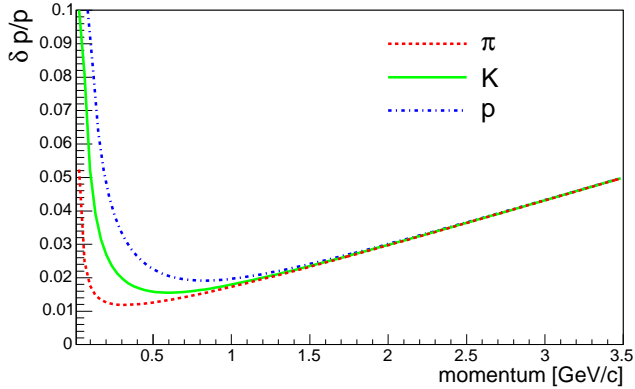}
\caption{Momentum resolution $\delta p/p$
calculated for the MRS $90^\circ$ setting
as a function of momentum with a magnetic field of 6~kG.
The curves are obtained from parametrization of the
mass resolution $\sigma^2_{m^2}$. 
At low momentum multiple scattering dominates the resolution 
while above 1~GeV the angular resolution of the spectrometers 
is the most important effect.
}
\label{fig:momres}
\end{figure}

The momentum resolution of the spectrometer 
can be inferred from the width of 
our mass-squared distributions since 
 $m^2=p^2[(ct/l)^2-1]$, where $c$ is the velocity of light,   
$l$ the track path length, and $t$ is the time-of-flight whose resolution
 is measured independently. 
The extracted resolutions were fit to the form 
$(\delta p/p)^2 = (c_{res}p)^2 + (c_{multi}/\beta)^2$, 
where $c_{res}$ is the contribution from the 
intrinsic angular resolution of the tracking detectors 
and $c_{multi}$ is the resolution from multiple scattering. 
The best fit is given by $c_{res} \sim 0.014~c$/GeV and
$c_{multi} \sim 0.01$ with D5 at 6~kG. 
Figure ~\ref{fig:momres} shows resolution curves based on this fit for
pions, kaons and protons when the D5 magnet is set to 6~kG. 
For the data presented in this paper the momentum resolution lies between
2 and 8\% depending on the momentum of the particle 
and the magnetic field in D5. 

\begin{figure}[htb]
\includegraphics[width=\columnwidth]{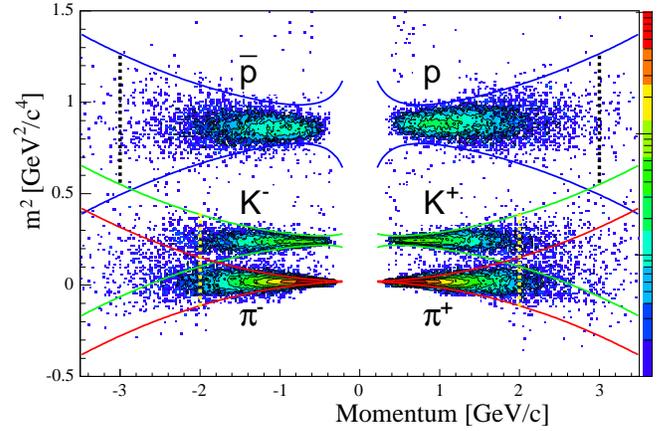}
\caption{Mass-squared distribution in TOFW as a function of 
reconstructed particle momentum obtained from tracking in TPCs 
through the D5 magnet at $y=0$. 
The curves in the figure are PID boundaries set at $\pm 3\sigma_{m^2}$
from the mean $m^{2}$ of the particle.
The dotted lines indicate the momentum cutoff for particle separation.}
\label{fig:mrspid}
\end{figure}

Particle Identification (PID) is based on time of flight data from the TOFW~ 
($\sigma_t \approx 80$ ps) with the start time taken from 
the BBC ($\sigma_t \approx 35$ ps). 
The TOFW is located at a distance of 4.3~m from the nominal interaction point. 
Final mass identification PID is based on cuts in the $m^2$ vs. $p$ space, 
as shown in Fig.~\ref{fig:mrspid}.
The cut boundaries are set at $\pm3\sigma$ from the mean $m^2$ values. 
TOFW provides $\pi/K$ separation to the momenta of 
1.85~GeV/$c$ and $K/p$ separation to the momenta of 2.85~GeV/$c$. 
For particle identification above these momenta,
the yields that are within the $\pm3\sigma$ overlapped regions 
are corrected for misidentified particles 
up to the momentum of 2~GeV/$c$ for $\pi/K$, 
and the momentum of 3~GeV/$c$ for $K/p$.
The kaon contaminations within the pion sample at 2~GeV/$c$ 
and within the proton sample at 3~GeV/$c$ are less than 1\%. 
The dotted lines in Fig.~\ref{fig:mrspid} show
the upper momentum limit used in the analysis on each particle type.
 
The data presented here were collected with the MRS at $90^\circ$,
for $y\sim0~(-0.1<y<0.1)$ and at $45^\circ-35^\circ$ for $y\sim1~(0.7<y<1.1)$
in RHIC Run II~(2001$-$2002).


From the number of identified particles, invariant differential yields are obtained 
from several spectrometer settings for different collision centralities. 
The invariant yields are corrected for geometrical acceptance 
and the efficiencies for detecting particles in the spectrometer. 
The inefficiencies arise from two effects,
inefficiencies due to single track loses and
those due to multiplicity dependent effects.

Geometrical acceptance factors are obtained from the GEANT~\cite{geant}
simulation package BRAG~(BRAHMS Analysis Geant), 
which is based upon the geometry and tracking capabilities 
of the the BRAHMS experimental setup.
The acceptance correction is calculated for each MRS setting 
and five different vertex windows covering the MRS track 
vertex range used in the analysis. 

\begin{figure}[htb]
\includegraphics[width=\columnwidth]{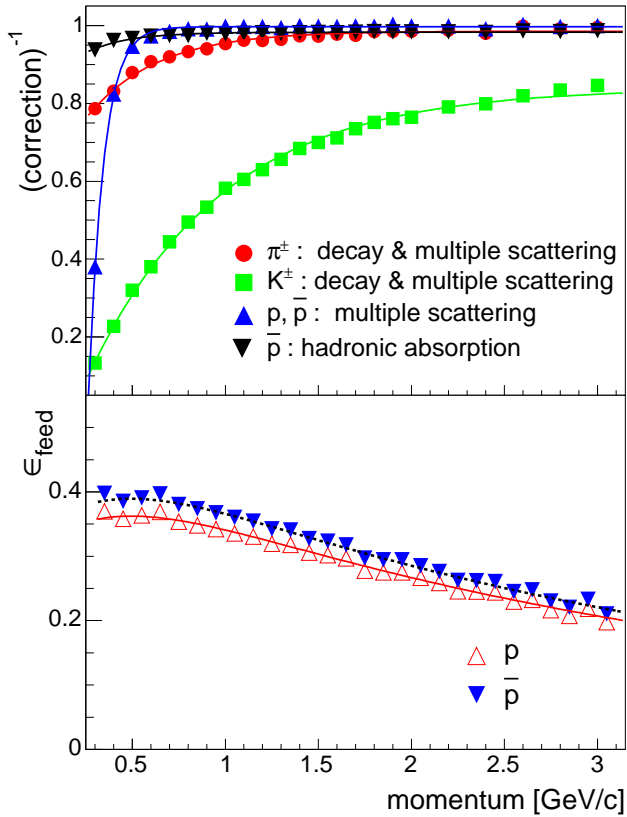}
\caption{Correction factors applied to the invariant yields, 
as discussed in the text.
The top panel shows the momentum dependence of the multiple scattering,
decay in flight and hadronic absorption effects. 
The bottom panel shows the fractional factors of $p$ and $\pbar$
from $\Lambda$ and $\bar{\Lambda}$ in all measured $p$ and $\pbar$,
$\epsilon_{feed}$, as a function of momentum.}
\label{fig:correction}
\end{figure}

The single track efficiency as a function of momentum in the spectrometer
is determined by a Monte Carlo simulation. Events with one particle are
first processed through BRAG with multiple scattering, decays, and
hadronic interactions processes included. 
In order to evaluate these effects, the simulated events 
are processed through the same digitization, 
reconstruction and particle selection algorithm 
that is applied to the real data. 

The upper panel of Fig.~\ref{fig:correction} shows correction factors
applied to the pion, kaon and proton spectra at y=0 to account 
for multiple scattering and ($\pi$ and K) decay in flight. 
The low momentum $\pbar$ spectra are also corrected 
for hadronic absorption in the beam pipe and detector materials. 
This effect amounts to $\sim 2-3$\% of the total yield. 
No corrections are applied for secondary protons,
arising mainly from the interaction of pions with the beam tube, 
since the contributions is found to be
negligible in MC simulations using HIJING as input, 
when the tracks were required to point back to the IP.
The difference of correction factors at different spectrometer settings
is below 1\%.

The multiplicity dependent track reconstruction efficiency has been studied 
by embedding simulated tracks into real events at the raw data level~\cite{truls}.
The combined events are reanalyzed to determine if the embedded tracks
are still reconstructed by the tracking program. 
Each track is associated with a number of pads in TPM1 and TPM2.
The resulting tracking efficiency is parameterized
as a function of the number of track-related pads found in the
two TPCs with signals above threshold for pions, kaons, and protons 
in various spectrometer angle settings. 
The mean number of track-related pad hits in the data sample 
varies from $\sim$350 to 60 as the centrality varies from 0\% to 60\%.
For the most central events, track reconstruction efficiency
is $\sim$85$-$95\% depending on spectrometer angle setting.

The efficiency for individual TOFW slats is investigated 
by projecting TPC tracks to the slat and comparing this 
to the distribution of TOFW hits.
The possibility of having multiple hits 
on a single slat is also corrected for.
The overall efficiency for particle 
identification using TOFW is estimated to be 90\%$\pm$2\% 
without significant dependences 
on spectrometer settings or collision centralities. 
The inefficiency includes uncertainties in matching tracks 
with hits in TOFW as well as the intrinsic detector inefficiencies.

Protons and antiprotons from weak decays lead to a contamination
of the primary hadron spectra. The proton and antiproton spectra
are corrected to remove the feed down contributions 
from $\Lambda$ and $\bar{\Lambda}$ weak decays. 
At midrapidity the ratio $N(\Lambda)=0.89N(p)$,  
$N(\bar{\Lambda})=0.95N(\bar{p})$ has been reported
in Au+Au collisions at $\snn$~=~130 GeV~\cite{star:lambda,phenix:lambda}.
We have studied the magnitude of the corrections using various
model assumptions as input to the BRAG simulations. 
Assuming primary $\Lambda$/$p$ ratios at $\snn$~=~200 GeV
similar to those measured at the lower energy~\cite{star:olga} 
and a constant behavior with rapidity,  
we take these ratios and measured spectra shapes as input to the BRAG code
for feed down correction from $\Lambda$ decays.
The simulated tracks are generated for the full phase space,
digitized, and go through the real data analysis algorithm, 
as is done to determine the other correction factors.
The lower panel of Fig.~\ref{fig:correction} shows 
the ratio~($\epsilon_{feed}$) of secondary $p$ and $\pbar$ 
to all measured $p$ and $\pbar$ as a function of momentum.
The fractional factors range from 25$-$40\%, and 
the largest value is $\sim$40\% around a momentum $p\sim 0.5$ GeV/$c$.
We multiply the proton and antiproton spectra by 1$-\epsilon_{feed}$
for all centrality windows and rapidities as a function of momentum.
The data are corrected on a track-by-track basis for efficiency
and feed down contributions.

\section{EXPERIMENTAL RESULTS}
\label{sec:results}

\begin{figure}[htb]
\includegraphics[width=\columnwidth]{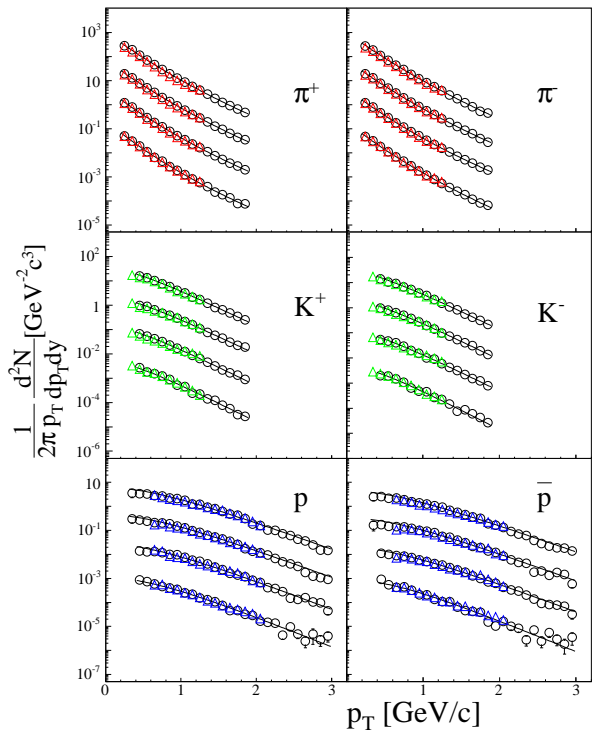}
\caption{The invariant yield spectra for the identified particles
for the 0$-$10\%, 10$-$20\%, 20$-$40\%, and 40$-$60\% 
central collisions in Au+Au collisions at $\snn$=200 GeV.
Circle(triangle) symbols are the results at $y=0$ ($y\sim1)$.
Curves overlaid to the data points are fits to the $y=0$ data,
as discussed in the text.
For clarity, the data points are scaled down by a factor of 10 successively 
from the top (0$-$10\%) to bottom(40$-$60\%) in decreasing order of centrality.
The error bars are statistical only.}
\label{fig:pt_spectra}
\end{figure}

Figure~\ref{fig:pt_spectra} shows the invariant spectra 
for charged hadrons, $\pi^{\pm}$, $K^{\pm}$, $p$, and $\bar{p}$ 
at different collision centralities.
The overlaid lines indicate fits to the data from $y=0$ in the range shown. 
The pion spectra are fitted with a power-law function, $A(1+p_T/p_0)^{-n}$. 
For kaons the spectra are best fit by an 
an exponential in $m_{T}$ , A[$e^{-m_{T}/T}$], where 
$m_{T} = \sqrt{p_{T}^2+m_{0}^2}$ where $m_{0}$ is the mass of the particle. 
The proton and antiproton $m_T$-spectra tend to deviate from
a single exponential shape, so a sum of two exponential 
functions is used to in the fit.
The point-to-point systematic uncertainties on the spectra 
and quality of fit are estimated 
by using other fit functions and varying the fit ranges.
The errors shown on the data points are statistical only.  
The overall systematic errors are estimated to be 10$-$15\%.
The main sources for the overall systematic errors are  
from the uncertainties in the normalizations 
used to calculate the invariant yields. 
Others are from uncertainties in estimating background contribution,
track reconstruction efficiencies, acceptance of spectrometer and
particle identification losses.
The yields and mean transverse momentum values are extracted from the fit functions.  
Tables~\ref{tab:fitrange} gives the fit ranges and the estimated percentage
of the total yield included in the fit ranges.

\begin{table}[htb]
\begin{ruledtabular}
\begin{tabular}{ccc}
	         &         $y=0$                 &       $y \sim 1$               \\ \hline
      $\pi^\pm$  & $0.2 < p_{T} < 1.9~(76.5\%)$   & $0.3 < p_{T}< 1.3~(72.1\%)$ \\
        $K^\pm$  & $0.4 < p_{T} < 1.9~(48.0\%)$   & $0.3 < p_{T}< 1.3~(40.9\%)$ \\
      $p, \pbar$ & $0.3 < p_{T} < 3.0~(72.1\%)$   & $0.6 < p_{T}< 2.1~(64.9\%)$  \\
\end{tabular}
\end{ruledtabular}  
\caption{Fit ranges for curves shown in Fig.~\ref{fig:pt_spectra}.
The yields were calculated from the data within the fit ranges, and
the estimated percentage is the ratio of measured yields within the fit ranges
to extrapolated yields for the full momentum range.}
\label{tab:fitrange}
\end{table}

\begin{figure}[htb]
\includegraphics[width=\columnwidth]{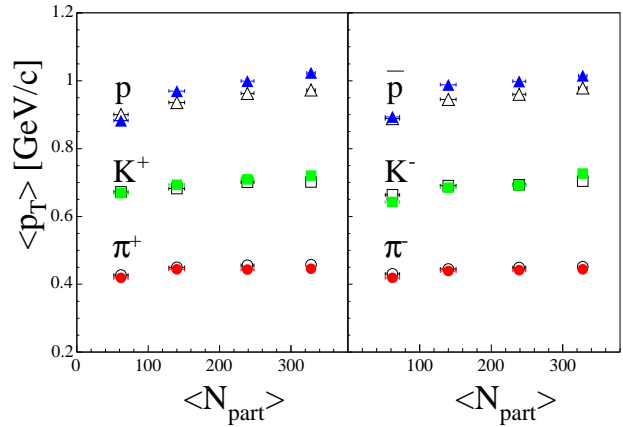}
\caption{
Mean $p_T$ extracted from the fit functions to the spectra 
as a function of centrality ($\npart$)
for $\pi^+$, $K^+$, $p$ (left) and  
$\pi^-$, $K^-$, $\bar{p}$ (right). 
Open symbols represent at $y=0$, and closed symbols are for $y\sim1$.
The error bars are statistical only.}
\label{fig:mean_pt}
\end{figure}

Figure~\ref{fig:mean_pt} shows the mean transverse momenta, $\meanpt$, 
as a function of $\npart$. We find that $\meanpt$ increases with
particle mass and centrality. This is suggestive of a hydrodynamic system
where the initial pressure increases with the number of participants.  

\begin{figure}[htb]
\includegraphics[width=\columnwidth]{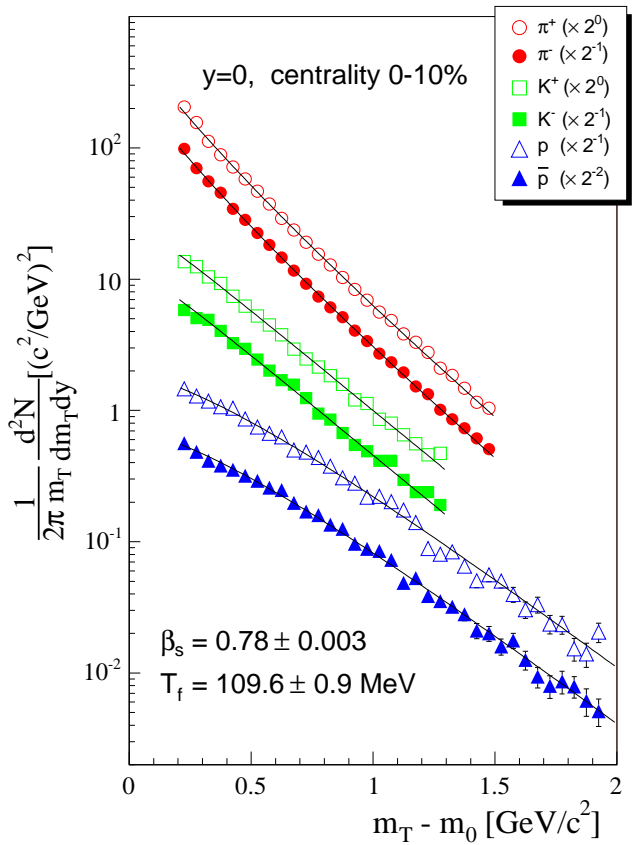}
\caption{The parametrization and the $\mt$ spectra for 0-10\% centrality.
The errors in the fit parameters are statistical only.}
\label{fig:bwfitAllspectra}
\end{figure}

\begin{figure}[htb]
\includegraphics[width=\columnwidth]{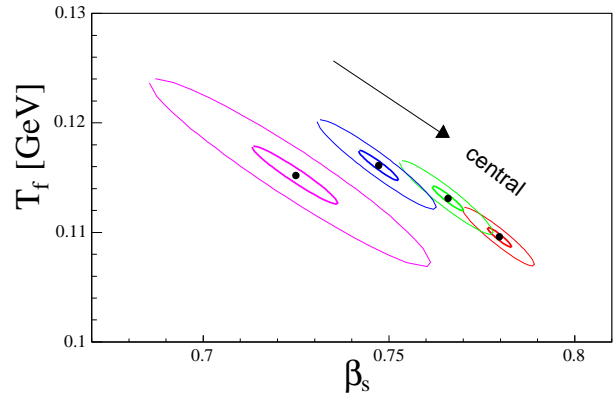}
\caption{Kinetic freeze-out temperatures $T_f$ 
and transverse flow velocities $\beta_s$ 
resulting from a simultaneous fit of
to $\pi^{\pm}$, $K^{\pm}$, $p$ and $\bar{p}$ spectra 
to Eq.~\protect{\ref{eq:blast-wave}}
as a function of collision centrality.
The curves correspond to increasing centrality as $\beta_s$ increases.
The points represent the best fit values of  $T_f$ and $\beta_s$ 
while the contours indicate 1$\sigma$ and 3$\sigma$ levels.
The systematic errors are $\le 5\%$ and are not included in the figure.}
\label{fig:flow}
\end{figure}

We also tried a hydrodynamic model fit to the experimental data with 
two free parameters: collective transverse flow velocity $\beta$ 
and the global thermal freeze-out temperature $T_{f}$.
We have utilized a version of a hydro-dynamically inspired 
``blast-wave" model initially developed to describe 
lower energy data ~\cite{ref:blast-wave}.
Assuming kinetic freeze-out of matter at constant $T_f$ 
with a collective transverse flow characterized 
by a velocity $\beta_s$, the invariant $m_T$ 
distribution can be described as follows:

\begin{equation}
\frac{dN}{m_{T}~dm_{T}} \propto  
\int_{0}^{R_{max}}~r~dr~m_{T} I_0({ p_{T}{\rm sinh}\rho \over T_f
})K_1({ m_{T}{\rm cosh}\rho \over T_f }),
\label{eq:blast-wave}
\end{equation}

where $T_f$ is the freeze-out temperature, $I_0$, $K_1$ are 
modified Bessel functions and 
$\rho={\rm tanh}^{-1}\beta_T$ is the transverse rapidity.
The transverse  velocity profile $\beta_T$ is parameterized by the
surface velocity $\beta_s$ : $\beta_{T}(r) = \beta_{s}(r/R_{max})^{\alpha}$.
Results of a simultaneous fit for 0$-$10\% centrality 
of Eq.~\ref{eq:blast-wave} to the spectra for $y=0$ 
is shown in Fig.~\ref{fig:bwfitAllspectra}. 
The parameters $\alpha$, $T_f$, and $\beta_s$ were allowed to vary,
as well as the normalization constants for each particle type.
The $p_{T}$ coverage of spectra at $y\sim1$ is not sufficient 
for a reliable hydrodynamic model fit.
The source parameter used was $R_{max}=13$ fm~\cite{ref:13fm}. 
The integral in ~Eq.~\ref{eq:blast-wave} is relatively insensitive to
changes in $R_{max}$, changing by less than 5\% 
when $R_{max}$ is changed from 5 to 20~fm.     

For the most central events (0$-$10\%), the fit yields values of 
$T_f$ $=$ 109.6$\pm$0.9 MeV, $\beta_s$ $=$ 0.78$\pm$0.003, 
and $\alpha$ $=$ 0.40$\pm$0.05. 
The average flow velocity $\langle \beta_{T} \rangle$ is then 
estimated to be 0.65 by taking an average 
over the transverse geometry~\cite{ref:beta}.
When fitting spectra from the other centrality windows 
we fixed the value of $\alpha$ to be 0.4.
The fits in the four centrality bins give a $\chi^2$/DOF between 1.34$-$1.49.
The systematic uncertainties in the fit parameters are estimated to be less than 5\%. 
Figure~\ref{fig:flow} shows the centrality dependence of the temperature and surface velocity.
$T_f$ decreases with centrality while $\beta_s$ increases. 
Since the surface velocity keeps increasing until the system decouples 
these results suggest that central collisions decouple later. 
The increased energy associated with the 
surface velocity requires a lower final temperature 
by energy conservation \cite{ref:flow_heinz}.

\begin{table}[htb]
  {\small
    \begin{tabular}{c c c c c}\hline\hline
       $y=0$     &  $0-10\%$             & $10-20\%$        & $20-40\%$        & $40-60\%$  \\ \hline
      $\pi^+$  & 283.3 $\pm$ 2.4    & 196.2 $\pm$ 1.9  & 119.7 $\pm$ 1.4  & 47.5 $\pm$ 0.96 \\
      $\pi^-$  & 277.9 $\pm$ 2.4    & 195.3 $\pm$ 2.0  & 118.8 $\pm$ 1.4  & 46.3 $\pm$ 0.89 \\
        $K^+$  &  45.0 $\pm$ 0.67   &  29.8 $\pm$ 0.54 &  17.9 $\pm$ 0.38 &  6.3 $\pm$ 0.26  \\
        $K^-$  &  40.9 $\pm$ 0.63   &  28.1 $\pm$ 0.52 &  16.3 $\pm$ 0.4  &  5.9 $\pm$ 0.25  \\
        $p  $  &  18.6 $\pm$ 0.21   &  13.4 $\pm$ 0.18 &   7.6 $\pm$ 0.13 &  2.81 $\pm$ 0.08  \\
       $\pbar$  & 13.7 $\pm$ 0.18   &   9.7 $\pm$ 0.2  &   5.6 $\pm$ 0.1  &  2.22 $\pm$ 0.08  \\\hline \hline
        $y \sim 1$       & $0-10\%$           & $10-20\%$        & $20-40\%$        & $40-60\%$  \\\hline
      $\pi^+$  & 266.8 $\pm$ 2.4    & 193.1 $\pm$ 2.2  & 117.6 $\pm$ 1.3  & 50.8 $\pm$ 1.01 \\
      $\pi^-$  & 276.2 $\pm$ 2.7    & 198.1 $\pm$ 2.5  & 120.6 $\pm$ 1.5  & 49.8 $\pm$ 1.0  \\
        $K^+$  &  40.6 $\pm$ 0.32   &  29.3 $\pm$ 0.27 &  17.2 $\pm$ 0.17 &  6.7 $\pm$ 0.12  \\
        $K^-$  &  38.5 $\pm$ 0.33   &  26.6 $\pm$ 0.29 &  16.1 $\pm$ 0.19 &  6.1 $\pm$ 0.13  \\
        $p  $  &  18.0 $\pm$ 0.17   &  12.7 $\pm$ 0.15 &   7.4 $\pm$ 0.1  &  2.97 $\pm$ 0.08  \\
      $\pbar$  &  12.1 $\pm$ 0.14   &   8.9 $\pm$ 0.13 &   5.5 $\pm$ 0.09 &  2.29 $\pm$ 0.08  \\\hline\hline
    \end{tabular}
  }
\caption{The yield $dN/dy$ from integration of extrapolated function
in each centrality bin at y=0 and $y\sim1$.
The fit range was shown in Tables~\ref{tab:fitrange}. 
The errors are statistical only.}
\label{tab:dndy_y0y1}
\end{table}

\begin{figure}[htb]
\includegraphics[width=\columnwidth]{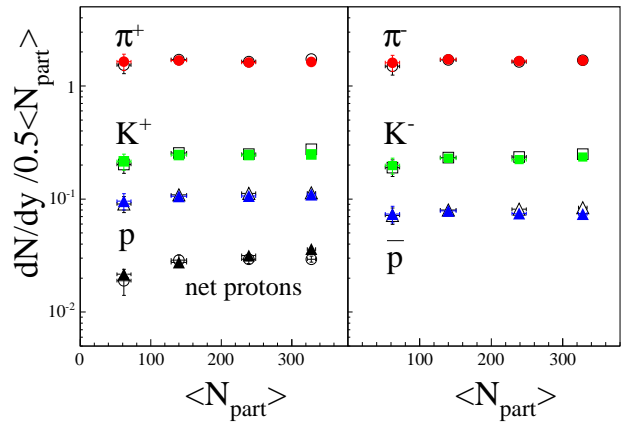}
\caption{ Rapidity density ($dN/dy$) per participant
pair ($N_{part}/2$) as a function of $N_{part}$ 
for $\pi^+$, $K^+$, $p$, net proton (left) and  
$\pi^-$, $K^-$, $\bar{p}$ (right). 
Open symbols represent at $y=0$, and closed symbols are for $y\sim1$.
The error bars represent the statistical errors only.}
\label{fig:dndy}
\end{figure}

The rapidity densities $dN/dy$ are determined for each particle 
by integrating the measured $\pt$ spectrum over $\pt$ 
using the previously discussed functional form 
to extrapolate outside of the region of the measurement. 
Table~\ref{tab:dndy_y0y1} shows the results at $y=0$ and $y\sim1$, respectively.
The $dN/dy$ values per participant pair are shown 
as a function of $\npart$ in Fig.~\ref{fig:dndy}. 
The systematic errors on the yields and $\meanpt$ values
from the extrapolation to the low momentum region 
are estimated as 5$-$10\%. 
The only centrality dependence evident is a small increase 
in the rapidity densities for the $K$ and $p$ channels 
in going to more central collisions.
The net proton rapidity densities show a increase 
from 40$-$60\% to 20$-$40\% and saturate after that. 
The proton excess,$(N_{p}-N_{\bar{p}})/(N_{p}+N_{\bar{p}})$, 
is 0.15$\pm$0.01$-$0.12$\pm$0.02 at $y=0$ and
0.19$\pm$0.01$-$0.13$\pm$0.02 at $y\sim1$ 
from peripheral to central collisions.
Our $\pbar/p$ ratio from $pp$ collisions showed 
a proton excess of 12\% at midrapidity~\cite{pp_run02}. 
This baryon asymmetry has been modeled at lower energy systems~\cite{dima96} 
where it has been found to be significantly greater~\cite{isr1, isr2}.

\begin{figure}[htb]
\includegraphics[width=\columnwidth]{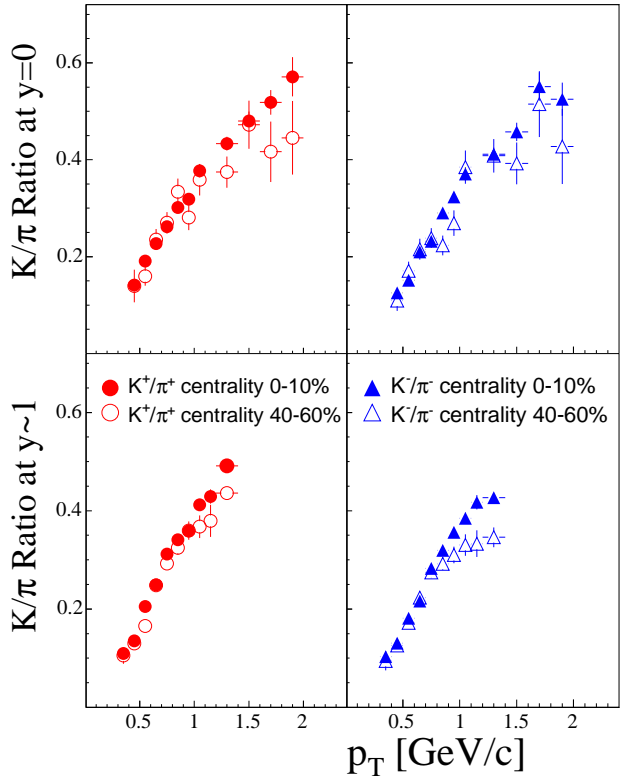}
\caption{$K^+/\pi^+$ (left) and $K^-/\pi^-$ (right) ratios 
as a function of $p_{T}$ at $y=0$ (top) and $y\sim1$ (bottom). 
Closed and open symbols represent central (0$-$10\%) and
peripheral (40$-$60\%) collisions, respectively.  
Only statistical error bars are shown.}
\label{fig:kpi_ratio}
\end{figure}

\begin{figure}[htb]
\includegraphics[width=\columnwidth]{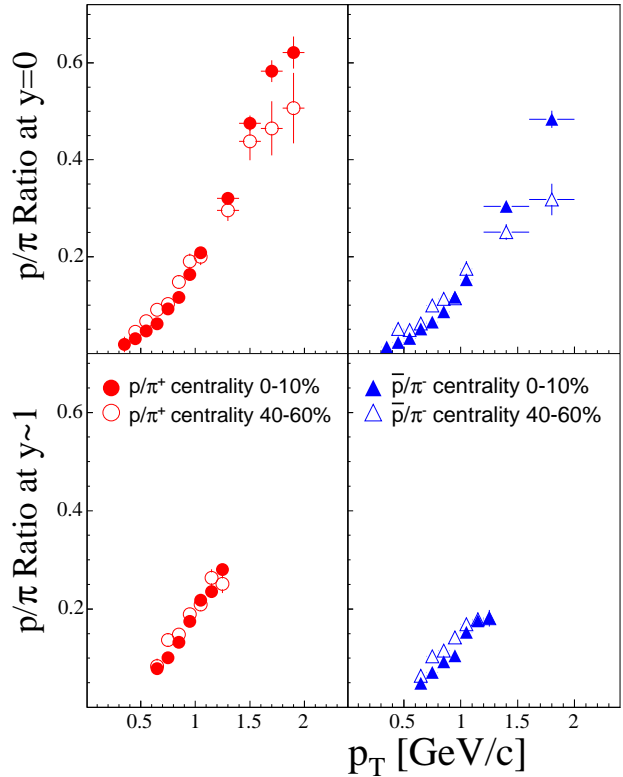}
\caption{$p/\pi^+$ (left) and $\bar{p}/\pi^-$ (right) ratios 
as a function of $p_{T}$ at $y=0$ (top) and $y\sim1$ (bottom). 
Closed and open symbols represent central (0$-$10\%) and 
peripheral (40$-$60\%) collisions, respectively.  
Only statistical error bars are shown.}
\label{fig:ppi_ratio}
\end{figure}

Figure~\ref{fig:kpi_ratio} compares the $K/\pi$ ratios 
as a function of $p_T$ for central and peripheral collisions.  
The ratios for both charges increase with $p_T$ at both rapidities. 
There is no significant centrality dependence below 1~GeV/$c$, however,
the increase is faster in central than peripheral collisions at higher momentum.
The weak centrality dependence of the $K/\pi$ production at RHIC~\cite{star130} 
differs from the measurements at lower energy 
from AGS~\cite{e8661} and SPS experiments~\cite{sps1}, 
where the $K^-/\pi^-$  ratio is enhanced ($\sim 1.5 - 2$) 
from central collisions compared to peripheral collisions. 
This might be attributed to the energy dependent longitudinal geometry 
of the colliding nuclei~\cite{fwang, brahmsMeson}, 
but further experimental and theoretical work
is needed to understand the observed energy dependence.

Figure~\ref{fig:ppi_ratio} shows the $p/\pi^{+}$ and $\pbar/\pi^{-}$ ratios
as a function of $p_{T}$ obtained at $y=0$ and $y \sim 1$ 
for central and peripheral collisions.  
For both centralities the ratios rise fast at low $p_{T}$.
Around $p_{T} \sim$ 2.0 GeV/$c$ the ratios increase from peripheral
to central collisions by 20\% for $p/\pi^{+}$ and by 50\% for $\pbar/\pi^{-}$ ratios.
Parton recombination and quark coalescence models describe qualitatively 
the observed baryon to meson ratios for central collisions
up through the intermediate $p_{T}$ region 
extending to $\sim$4$-$5~GeV/$c$~\cite{hwa, greco}.

\begin{figure}[htb]
\includegraphics[width=\columnwidth]{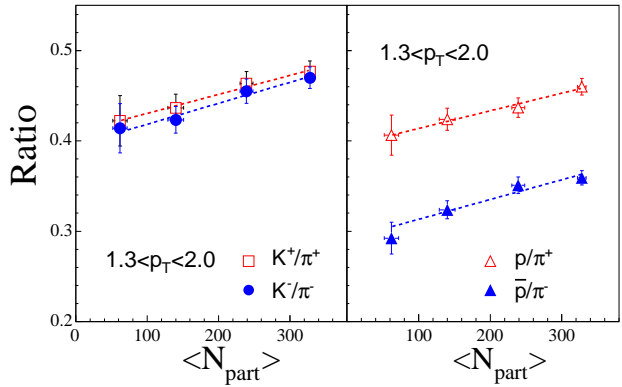}
\caption{The ratio of $K/\pi$ (left) and $p/\pi$ (right) 
as a function of $N_{part}$ for $1.3<p_{T}<2.0$ GeV/$c$ at mid-rapidity.
Open and closed symbols represent the ratio of positive particles and 
negative ones, respectively.}
\label{fig:highpt_ratio}
\end{figure}

To clarify the centrality dependence of particle ratios at higher $p_{T}$, 
and to see if the ratios of high $\pt$ particles 
are sensitive to the size of the interaction volume,
we present the $K/\pi$ and $p/\pi$ ratios versus $N_{part}$.
Figure~\ref{fig:highpt_ratio} shows that the $K/\pi$ and $p/\pi$ ratios for
$1.3<p_{T}<2.0$ GeV/$c$ increase with $N_{part}$. 
This increase is similar for protons and kaons
with little difference in slope 
between the particle and antiparticle ratios. 
The behavior suggests that rescattering and/or 
hydrodynamic effects are stronger for larger collision volumes.

\section{SUMMARY}
\label{sec:summary}
 
Particle production of identified charged hadrons, 
$\pi^{\pm}$, $K^{\pm}$, $p$, and $\bar{p}$
in Au+Au collisions at $\snn =$ 200 GeV
has been studied as a function of transverse momentum and collision centrality 
at $y=0$ and $y\sim1$ by the BRAHMS collaboration at RHIC. 
Significant collective transverse flow at kinetic freeze-out 
is estimated for both central and mid-central events. 
The magnitude of the radial expansion increases 
with the collision centrality indicating more hydro-like collectivity 
in the transverse direction for the central collisions.
The $p$, $\pbar$ and $K^{\pm}$ yields relative to the 
pion production at RHIC show a strong transverse momentum dependence. 
Contrary to lower energy results no significant 
centrality dependent $K/\pi$ ratios below 1~GeV/$c$ 
are observed at the RHIC energy.
The $p/\pi^{+}$ and $\pbar/\pi^{-}$ ratios 
increase with $p_{T}$ and centrality 
reaching a value of 0.6 and 0.5 at $p_T \sim 2~$GeV/$c$.  
The particle yield scaled by  $N_{part}$ is nearly constant,
and only weakly increasing with centrality for all particles.
No significant changes for the bulk properties in hadron production
are observed within one unit around midrapidity in Au+Au collisions
at $\snn$~=~200~GeV. 

\begin{acknowledgments}

This work was supported by the division of Nuclear Physics of the
Office of Science of the U.S. DOE, the Danish Natural Science
Research Council, the Research Council of Norway, the Polish State
Committee for Scientific Research and the Romanian Ministry 
of Education and Research. We thank the staff of the Collider-Accelerator 
Division of BNL for their excellent and dedicated work to deploy RHIC
and their support to the experiment.

\end{acknowledgments}

{}

\end{document}